\documentclass[aps,amsmath,amssymb,nofootinbib,superscriptaddress,showpacs,floatfix,prb,twocolumn]{revtex4-1}

\usepackage{latexsym}
\usepackage{graphicx}
\usepackage{times,psfrag}
\usepackage{amsmath}
\usepackage{dcolumn}
\usepackage{bm,bbm}       
\usepackage{color}
\usepackage{latexsym,amsmath,amssymb,bm,euscript}
\usepackage{color}

\renewcommand\Re{\operatorname{Re}}  
\renewcommand\Im{\operatorname{Im}}

\begin{document}

\title{Structure of vortex-bound states in spin-singlet chiral superconductors}

\author{Darrick Lee}
\affiliation{Max-Planck-Institut f\"ur Festk\"orperforschung, Heisenbergstrasse 1, D-70569 Stuttgart, Germany}
\affiliation{Quantum Matter Institute, University of British Columbia, Vancouver BC, Canada V6T 1Z4} 

\author{Andreas P. Schnyder}
\email{a.schnyder@fkf.mpg.de}
\affiliation{Max-Planck-Institut f\"ur Festk\"orperforschung, Heisenbergstrasse 1, D-70569 Stuttgart, Germany}

\date{\today}

\begin{abstract}
We investigate the structure of vortex-bound states in spin-singlet chiral  superconductors with ($d_{x^2-y^2} \pm i d_{xy}$)-wave and 
($d_{xz} \pm id_{yz}$)-wave pairing symmetries.  
It is found that vortices in the \mbox{($d_{xz} \pm id_{yz}$)}-wave state bind
zero-energy states which are dispersionless along the vortex line, forming a doubly degenerate Majorana flat band.
Vortex-bound states of ($d_{x^2-y^2} \pm i d_{xy}$)-wave superconductors, on the other hand, exist
only at finite energy. Using exact diagonalization and analytical solutions of tight-binding Bogoliubov-de Gennes Hamiltonians, we compute the energy spectrum of the vortex-bound states and
the local density of states around the vortex and antivortex cores.  
We find that the tunneling conductance peak of the vortex is considerably broader than 
that of the antivortex. This difference can be used as a direct signature of the chiral order parameter symmetry.
\end{abstract}

\maketitle

\section{Introduction}
Chiral superconductors are attracting growing interest because of their potential use for
novel superconducting devices and quantum information technology.
These unconventional superconductors exhibit pairing gaps whose phase winds around the Fermi surface
in multiples of $2\pi$, leading to a non-trivial wave function topology and a breaking of time-reversal symmetry.
The non-trivial topology gives rise to a multitude of interesting phenomena~\cite{sigristRMP91,Schnyder08,ryuNJP10,black_schaffer_review,chiu_review15,schnyder_review15,hasan:rmp,qi:rmp}, in particular 
 subgap states in vortex cores and protected gapless edge modes that can carry
quantized thermal current and particle current. 
Probably the most prominent example of a chiral superconductor is the spin-triplet ($p_x \pm i p_y$)-wave state, which 
is believed to be realized in Sr$_2$RuO$_4$~\cite{MaenoRMP03,maenoJPSJ12}, in the A phase of superfluid ${}^3$He~\cite{Volovik1988,volovkiFlatBandReview,volovikLectNotes13},
and in two-dimensional cold atomic gases~\cite{Gurarie20072}. 
Spin-polarized ($p_x \pm i p_y$)-wave superconductors 
support non-degenerate Majorana zero-en\-ergy modes localized at vortex cores~\cite{readGreen,KopninSalomaa91,volovikJETP99,krausPRL08,krausPRB09,moellerPRB11}.
These Majorana quasiparticles obey non-Abelian statistics 
and can therefore be employed to implement  topological quantum computing~\cite{Kitaev20032,nayakRMP08}.

Another example of a chiral superconductor is the spin-singlet chiral $d$-wave state~\cite{black_schaffer_review}. 
The non-trivial topology of this phase is analogous to that of the chiral  ($p_x \pm i p_y$)-wave superconductor.
However, due to the conservation of spin-rotation symmetry, the  edge modes
of spin-singlet chiral superconductors carry besides a thermal current also a well-defined quantized spin current.
Recently, it has been proposed that graphene doped to the van Hove filling is a potential
experimental realization of the spin-singlet chiral superconductor~\cite{chubukovNatPhys}.
Other candidate materials for spin-singlet superconductivity with broken time-reversal symmetry  include 
\mbox{SrPtAs}~\cite{youn2011,youn2012,Biswas2013,Fischer2014}, the heavy fermion system URu$_2$Si$_2$~\cite{goswamiArXiv13,Hsu14,Kasahara07,Li13,Sumiyoshi14,Schemm2014,Mydosh2011},
and Cu-doped TiSe$_2$~\cite{morosanNatPhys06,tailleferPRL07,ganeshEfremov2014}.

In this paper, we investigate the energy spectrum and the wave function profile of vortex-bound states in spin-singlet chiral superconductors
with ($d_{x^2-y^2} \pm i d_{xy}$)-wave and 
($d_{xz} \pm id_{yz}$)-wave pairing symmetries.
Interestingly, we find that vortices in the ($d_{xz} \pm id_{yz}$)-wave state 
support zero-energy states with a flat dispersion along the vortex line (Fig.~\ref{fig:spectrum}).
The  ($d_{x^2-y^2} \pm i d_{xy}$)-wave state, on the other hand, supports
vortex bound states only at finite energy.  We show that for both pairing symmetries
the tunneling conductance peak of the vortex is about twice as broad as that of the antivortex (Figs.~\ref{fig:cldosx2y2} and~\ref{fig:cldosxz}).
This property is present even at temperatures considerably higher than the energy spacing between the vortex-bound states, and can be used as a direct probe
of time-reversal symmetry breaking and chiral order parameter symmetry.

The remainder of the paper is structured as follows. In Sec.~\ref{sec:model} we introduce the Bogoliubov-de Gennes (BdG) Hamiltonian of the
spin-singlet chiral superconductors in the presence of a vortex/antivortex pair. The analytical solutions of the vortex-bound state
wave functions are derived in Sec.~\ref{analytical}. In Sec.~\ref{numerical}
we present the numerical results for the local density of states 
around the vortex and antivortex cores and discuss the asymmetry between 
 vortex and antivortex bound states. Our conclusions and a  discussion of  implications for experiments are given in Sec.~\ref{sec:summary}.
 Some technical details 
 of the derivation of the vortex-bound states are provided in 
 Appendix~\ref{firstAppendix}.

\section{Bogoliubov-de Gennes Theory}
\label{sec:model}

\begin{figure} [t!]
\includegraphics[width=0.45\textwidth]{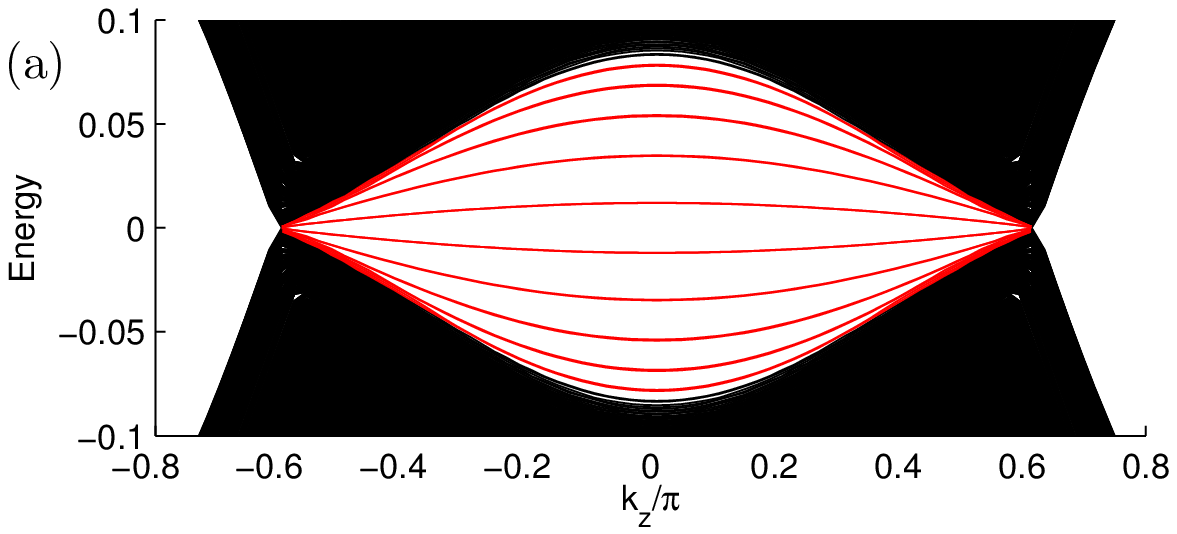}
\includegraphics[width=0.45\textwidth]{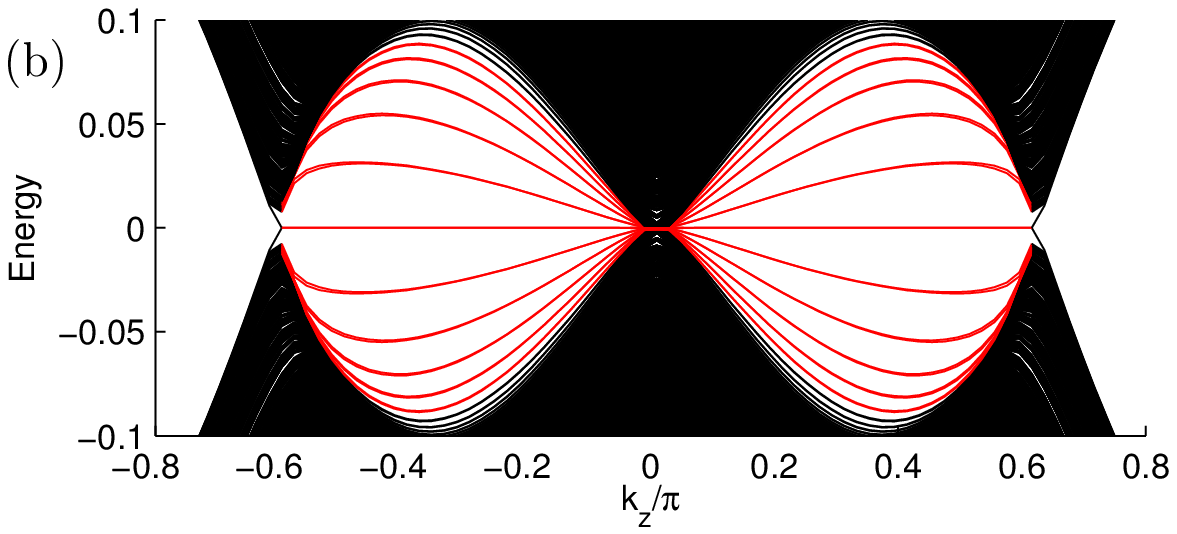}
\caption{(Color online) Energy Spectrum of (a) the $(d_{x^2-y^2}+id_{xy})$-wave and (b) the $(d_{xz}+id_{yz})$-wave pairing state in the presence of a vortex/antivortex pair. The vortex-bound states are highlighted in red. }
\label{fig:spectrum}
\end{figure}

At a phenomenological level chiral $d$-wave superconductors can be described by the $2 \times 2$ BdG Hamiltonian 
 $\mathcal{H}=\frac{1}{2}\sum_{\bf k}\Phi^\dag_{\bf k}H_{\bf k}\Phi^{\phantom{\dag}}_{\bf k}$, with 
\begin{eqnarray}  \label{Ham} 
H_{\bf{k} }
=
\begin{pmatrix}
h_{\bf k}&  
\Delta_{\bf k}  \cr
\Delta^{\dag}_{\bf k}  & -h_{\bf -k}^{\mathrm{T}} \cr   
\end{pmatrix} 
\end{eqnarray}
and the Nambu spinor $\Phi_{\bf k} = ( c^{\phantom{\dag}}_{{\bf k} \uparrow} , c^{\dag}_{-{\bf k} \downarrow}  )^{\mathrm{T}}$.
Here, $c^{\dag}_{{\bf k} s}$ ($c^{\phantom{\dag}}_{{\bf k} s}$)
represents the electron creation (annihilation) operator with momentum ${\bf k}$ and spin $s$.
The normal state $h_{\bf k} = t ( \cos k_x + \cos k_y) + t_z \cos k_z - \mu$ 
describes electrons hopping between nearest neighbor sites of a tetragonal lattice, 
where $t$ and $t_z$ denote the hopping integrals in the $xy$ plane and along the $z$ axis, respectively,
and $\mu$ is the chemical potential. 
In the following  we focus on quasi-two-dimensional systems with $t_z \ll t$ and consider two different spin-singlet chiral paired states,
namely the ($d_{x^2-y^2} \pm i d_{xy}$)-wave state described by
\begin{subequations}
\begin{equation}
\label{eq:gapx2y2}
\Delta_{\bf k} 
=
\Delta_0 
(\cos k_z + 4) \left( \cos k_x - \cos k_y \pm i \sin k_x \sin k_y \right) \; \; \;
\end{equation}
and the ($d_{xz} \pm id_{yz}$)-wave state give by
\begin{eqnarray}
\label{eq:gapxz}
\Delta_{\bf k} 
= 
\Delta_0
 \left(
\sin k_x \sin k_z \pm i \sin k_y \sin k_z
\right) ,
\end{eqnarray}
\end{subequations}
where $\Delta_0$ denotes the superconducting gap energy.
The superconducting order parameter for both pairing symmetries 
exhibits point nodes at the north and south poles of the Fermi spheroid.
The gap function~\eqref{eq:gapxz} has in addition a line node at
the equator of the Fermi surface, see Fig.~\ref{fig:spectrum}.
The point nodes of the ($d_{x^2-y^2} \pm i d_{xy}$)-wave state~\eqref{eq:gapx2y2} realize double Weyl nodes, whose
stability is protected by a Chern number that takes on the values $\pm 2$.\cite{schnyder_review15}
The low-energy nodal quasiparticles near these double Weyl nodes exhibit linear and quadratic dispersions along the $k_z$ direction and in the $k_x k_y$ plane, respectively. This anisotropic dispersion leads to a density of states which increases linearly with energy.
The point nodes of the  ($d_{xz} \pm id_{yz}$)-wave state~\eqref{eq:gapxz}, on the other hand, correspond to single Weyl nodes with Chern number $\pm 1$~\cite{goswamiArXiv13}.

According to the classification of Ref.~\onlinecite{chiu_review15}, $H_{\bf{k} }$ belongs to symmetry class C, since it satisfies 
particle-hole symmetry
\begin{eqnarray}
C H_{\bf{k} } C^{-1}
= - H_{- \bf{k} }, 
\end{eqnarray}
with $C=\sigma_x \mathcal{K}$ and $C^2 = - \mathbbm{1}$, but breaks time-reversal symmetry.
In the following we consider vortex lines along the $z$ axis, which allows us to make use of the translation symmetry along the $z$ direction.
Therefore, we can decompose  $H_{\bf{k} }$ into a family of two-dimensional layers with
fixed $k_z$. 
Hence, the analysis of vortex-bound states of the three-dimensional superconductor~\eqref{Ham}, reduces to the
problem of studying point vortices of two-dimensional superconductors as a function of $k_z$.
By the bulk-defect correspondence of Refs.~\onlinecite{ryuNJP10,chiu_review15,TeoPRB10,Freedman2011}, it follows that a two-dimensional Hamiltonian in symmetry class C
does not exhibit any zero-energy vortex-bound states. This is the case for the  ($d_{x^2-y^2} \pm i d_{xy}$)-wave pairing state.
The superconductor with  ($d_{xz} \pm id_{yz}$)-wave gap symmetry, however, constitutes an exception to this rule.
This is because for fixed $k_z$ Eq.~\eqref{eq:gapxz} does not have
 chiral $d$-wave symmetry, 
rather it exhibits a chiral $p$-wave character and thus belongs to symmetry class D. 
As a consequence, we find from the classifications of Refs.~\onlinecite{ryuNJP10,chiu_review15,TeoPRB10,Freedman2011} that 
vortices in the  ($d_{xz} \pm id_{yz}$)-wave state
support zero-energy  bound-states
protected by a Chern-Simons invariant (see Fig.~\ref{fig:spectrum}) \cite{volovkBulkVortex}.

\subsection{Implementation of Vortex/Antivortex Pair}

In the following we discuss how vortex lines along the $z$ axis 
are implemented on a microscopic level.
As mentioned above, due to translation symmetry along $z$ we can decompose the three-dimensional Bogoliubov equations into
a family of two-dimensional equations parametrized by $k_z$.
In order to introduce   vortex/antivortex pairs we 
Fourier transform   
the $k_x$ and $k_y$ components of Eq.~\eqref{Ham} into real space.
This yields a two-dimensional lattice Hamiltonian on an $N \times N$ square lattice, 
with coordinates ranging from $[-N/2,N/2)$. 
In order to suppress possible surface states of the superconductor we impose
closed boundary conditions in all  directions.
The vortex/antivortex pair is implemented by applying a radial profile $h(r)$ and phase $\phi(x,y)$ to the gap parameter,
\begin{subequations} \label{defVortexAntivortex}
\begin{eqnarray}
\Delta_0 \rightarrow \Delta_0 h^2(r) e^{in \phi(x,y)},
\end{eqnarray}
where $n$ is the vorticity of the vortex/antivortex pair. 
Note that the vortex (antivortex) is defined by a positive (negative) winding number $\oint_\mathcal{C} \arg[\Delta({\bf r})]\, ds$ about the vortex (antivortex) core, 
where $\mathcal{C}$ is a small circle centered at the core.
The function that parametrizes the phase of the vortex/antivortex pair is given by~\cite{ChangNCSvortex}
\begin{eqnarray} \label{PhiFun}
\phi (x,y) = \tan^{-1}\left(\frac{2ayC(y)}{x^2 + y^2 - a^2}\right),
\end{eqnarray}
where $a$ is half the distance between the vortex and the antivortex. 
To minimize finite-size effects we choose $a=N/4$, such that the distance between
the vortex and the antivortex is maximized. With this choice, the vortex and antivortex
behave like isolated vortices for $N$ large enough.
The factor $C(y) = \left|1+A-\frac{2|y|}{N}\right|^{1/B}$ in Eq.~\eqref{PhiFun} is used to provide continuity of the gap phase across the closed boundary, while retaining the structure of the original gap phase around the vortex cores. The values $A$ and $B$ are determined by an optimization process~\cite{footnote1}.
The radial profile of the vortex and antivortex at $(a, 0)$ and $(-a, 0)$, respectively, is taken to be
\begin{align}
	h(r) = \left\{ \begin{array}{cl}
		0 & : 0 \leq r < 1 \\
		\sqrt{\tanh(r/\rho)} & :r \geq 1
	\end{array}
	\right. ,
\end{align}
\end{subequations}
where $\rho$ is the size of the (anti)vortex and $r$ the distance from the (anti)vortex core.
 The piecewise nature of the profile is used to remove unphysical singularities at the vortex core.
We observe that the profile  $h(r)$ is linear close to the core and  constant far away from the core.

\section{Structure of vortex-bound states}
\label{sec:zwei}

In this section, we study the structure of the vortex-bound states using both analytical and numerical methods.
For the analytical solutions of the vortex-bound states we focus on the $(d_{x^2-y^2}+id_{xy})$-wave pairing state. 
The vortex-bound state wavefunctions of the $(d_{xz}+id_{yz})$-wave superconductor can be inferred
from the published results on vortex-bound states of the chiral $(p_x+ip_y)$-wave state~\cite{KopninSalomaa91,volovikJETP99,moellerPRB11}. That is, the
vortex-bound states of the $(d_{xz}+id_{yz})$-wave state~\eqref{eq:gapxz}, are obtained from the bound-state solutions of Refs.~\onlinecite{KopninSalomaa91,volovikJETP99,moellerPRB11} by scaling 
the gap energy by  $\sin(k_z)$ [i.e., $\Delta_0 \to \Delta_0\sin(k_z)$].

\subsection{Analytical solutions for $(d_{x^2-y^2}+id_{xy})$-wave state}
\label{analytical}

\begin{figure} [t!]
\includegraphics[width=0.23\textwidth]{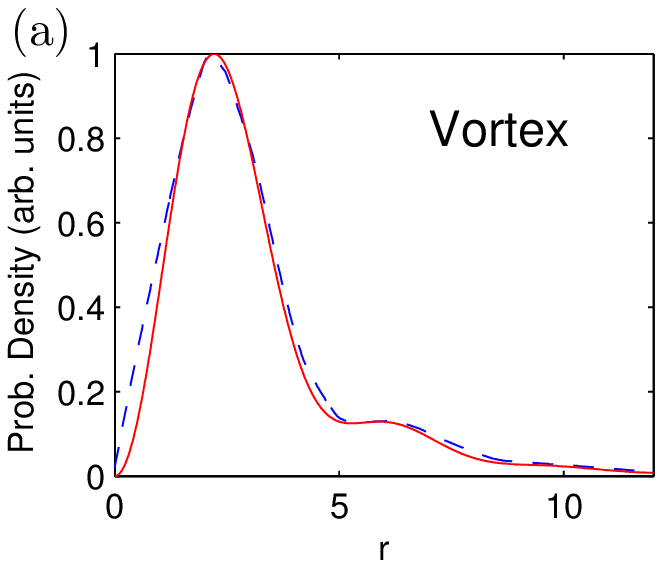}
\includegraphics[width=0.23\textwidth]{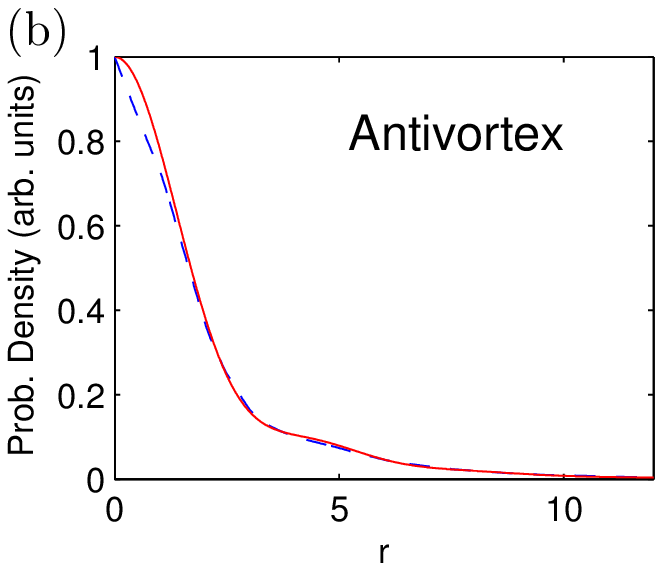}
\includegraphics[width=0.23\textwidth]{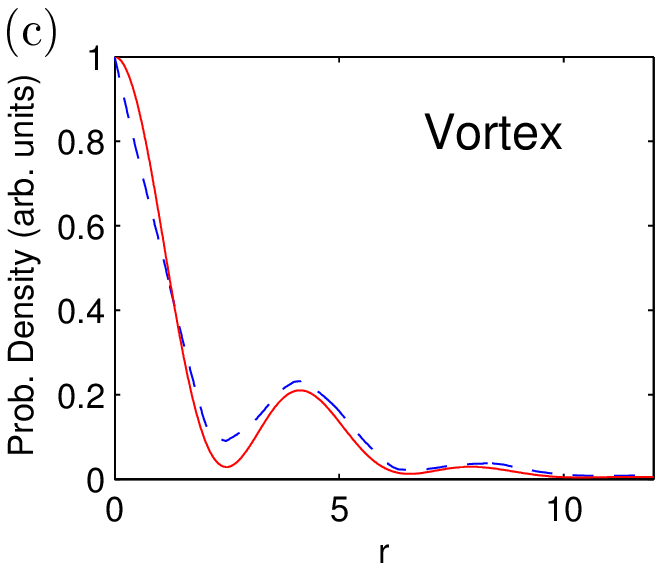}
\includegraphics[width=0.23\textwidth]{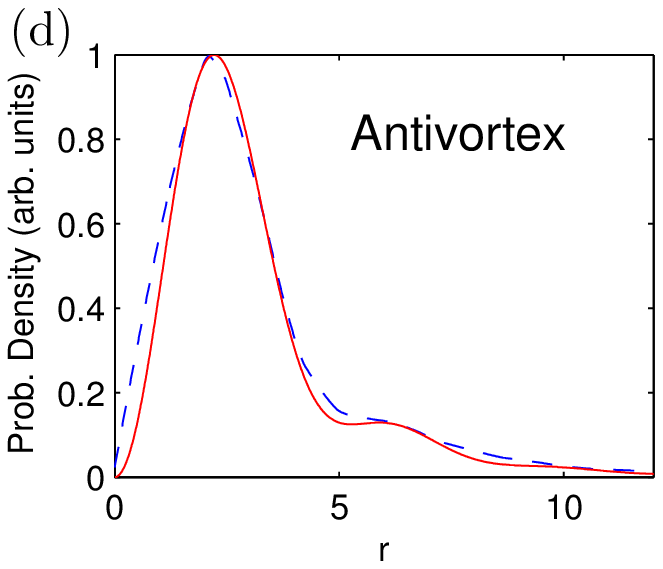}
\caption{(Color online) Comparison between 
the analytical (solid red) and numerical (dashed blue) solutions of the (anti)vortex-bound state wavefunctions of the $(d_{x^2-y^2}+id_{xy})$-wave superconductor.
(a), (b) Wavefunction amplitude of the lowest energy vortex- and antivortex bound state, respectively.
(c), (d) Wavefunction amplitude of the second lowest energy vortex and antivortex-bound state, respectively.}
\label{fig:ancomp}
\end{figure}

In order to obtain analytical expressions for the vortex-bound states 
of the $(d_{x^2-y^2}+id_{xy})$-wave state,
we first derive a low-energy continuum description of Hamiltonian~\eqref{Ham}.
To this end, we consider a single (anti)vortex at the origin and assume that the Fermi surface is small, of spherical shape, and centered at the $\Gamma$ point.
Performing a small momentum expansion, we obtain for the normal state
 $h({\bf k}) = \frac{1}{2m}(k_x^2-k_y^2) + \frac{1}{2m_z} k_z^2- \mu$ and the gap function~\cite{footnote2}
$\Delta({\bf k}) = \Delta_0[(1-k_z^2)+4](k_x^2 - k_y^2 + 2ik_x k_y)$. 
Since we consider a quasi-two-dimensional system, we can take $k_z$ to be a fixed parameter 
and absorb all $k_z$ dependent terms in constants. That is, we let 
 $\mu - \frac{1}{2m_z}{k_z^2}\rightarrow \mu$ and $ \Delta_0[(1-k_z^2)+4]\rightarrow \Delta_0$. 
By replacing momentum variables by momentum operators, i.e.,  $(k_x, k_y) \rightarrow -i(\partial_x,\partial_y)$, we arrive at the real-space representation of the continuum Bogoliubov equations
\begin{eqnarray}  \label{HamCont} 
\begin{pmatrix}
\hat{h}&  
\hat{\Delta} \cr
\hat{\Delta}^{\dag}  & -\hat{h}^{\mathrm{T}} \cr   
\end{pmatrix} 
\begin{pmatrix}
u \cr
v
\end{pmatrix}
= \epsilon
\begin{pmatrix}
u \cr
v
\end{pmatrix},
\end{eqnarray}
where
\begin{align}
\hat{h} &= -\frac{\nabla^2}{2m} - \mu,   
\cr 
\hat{\Delta} &=\sqrt{\Delta({\bf r})} (-\partial_x^2+\partial_y^2 - 2i\partial_x \partial_y)\sqrt{\Delta({\bf r})} ,
\end{align}
and $\Delta({\bf r}) = \Delta_0 h^2(r) e^{in \theta}$ describes an (anti)vortex at the origin with
vorticity $n$ and polar coordinates $(r, \theta)$.
To simplify the analysis of the Bogoliubov equations we rescale the equations in terms of
the characteristic length
\begin{eqnarray}
L &= \frac{1}{\sqrt{\Delta_0^2 m^3 \mu}} ,
\end{eqnarray}
which yields the dimensionless variables
\begin{eqnarray}
\overline{x} & = \frac{x}{L} \quad \textrm{and} \quad
\overline{\epsilon} =  \epsilon m L^2 = \frac{\epsilon}{\Delta_0^2 m^2 \mu} .
\end{eqnarray}
With this, the dimensionless vortex profile becomes
\begin{align}
	h(\overline{r}) = \left\{ \begin{array}{cl}
		0 & : 0 \leq \overline{r} < 1/L \\
		\sqrt{\tanh(L\overline{r}/\rho)} & : \overline{r} \geq 1/L
	\end{array} .
	\right.
\end{align}
For ease of notation, we omit the overbars for the remainder of this section. I.e., in the following all
variables are assumed to be dimensionless.  
By setting
 $u({\bf r}) = \exp[i(l + \frac{n+2}{2})\theta] u(r)$ and $v({\bf r}) = \exp[i(l - \frac{n+2}{2})\theta] v(r)$, where $l = \frac{2\mathrm{k}-1}{2}$ is 
 restricted to half integers ($\mathrm{k} \in \mathbb{N}$),
 we obtain for the Bogoliubov equations  
\begin{subequations} \label{eq:bdgdl}
\begin{eqnarray} 
\left[ -\frac{1}{2} L_{M_+} - \gamma^2\right]u + \frac{1}{\gamma}D_+ v &= \epsilon u , \\
\left[ \frac{1}{2} L_{M_-} + \gamma^2\right]v + \frac{1}{\gamma}D_- u &= \epsilon v ,
\end{eqnarray}
with $\gamma = 1 / ( \Delta_0 m) $, $M_\pm = \frac{1}{2}(2\pm 2l + n)$, and the second order differential operators 
\begin{eqnarray}
L_s &= \partial_r^2 + \frac{1}{r}\partial_r - \frac{s^2}{r^2} 
\end{eqnarray}
and  
\begin{eqnarray}
D_\pm = \biggl[&\left(\frac{h^2(1-l^2)}{r^2} + \frac{hh'(-1\pm 2l)}{r} - hh''\right) \cr
&+ \left(\frac{h^2(-1\pm 2l)}{r} - 2hh'\right)\partial_r - h^2\partial_r^2 \biggr] .
\end{eqnarray}
\end{subequations}

\begin{figure} [t!]
\includegraphics[width=0.45\textwidth]{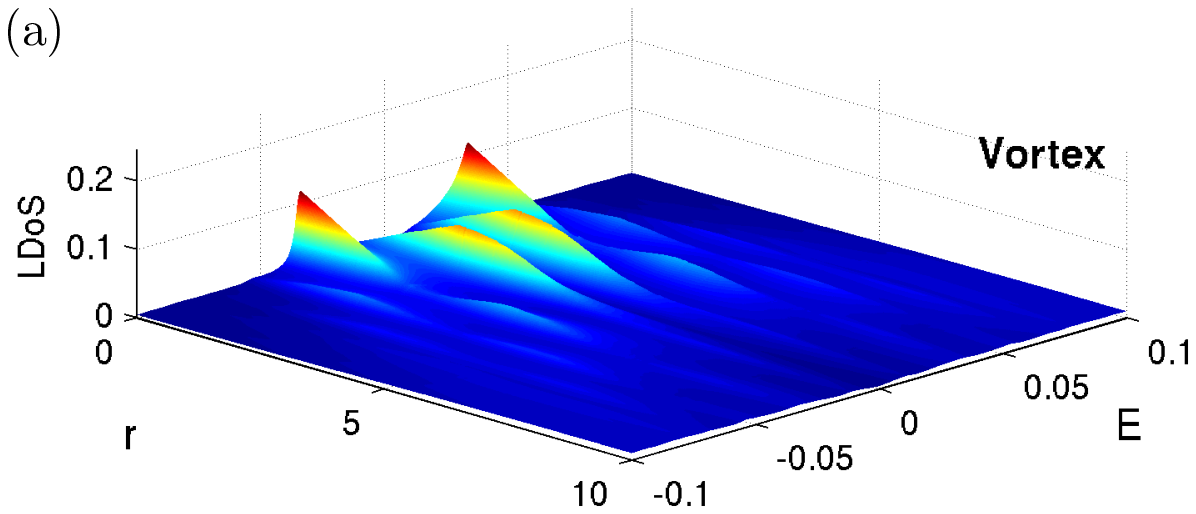}
\includegraphics[width=0.45\textwidth]{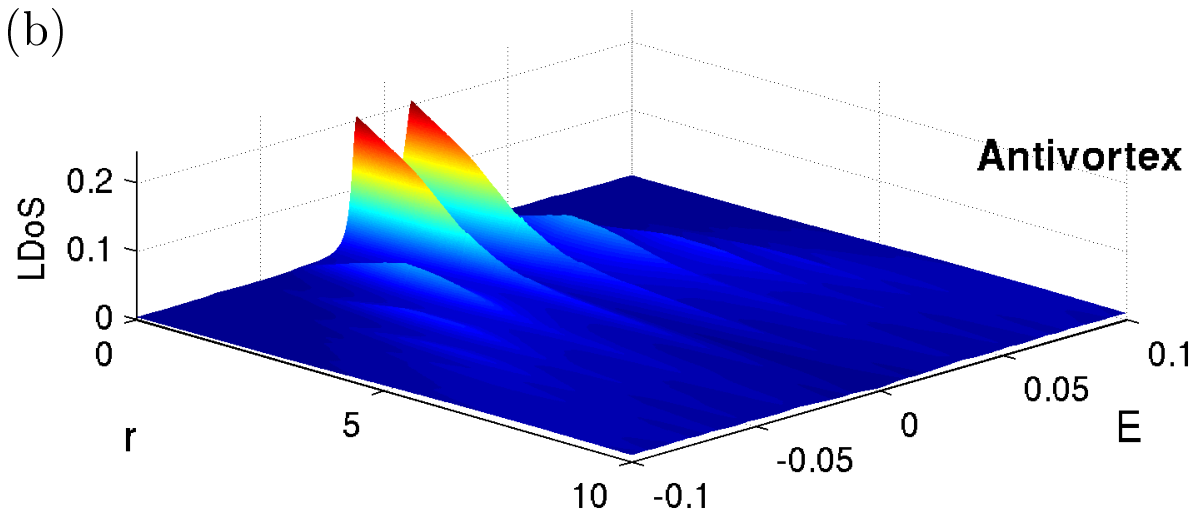}
\caption{(Color online) Local density of states of Eq.~\eqref{defLDOS} near (a) the vortex and (b) the antivortex of the $(d_{x^2-y^2}+id_{xy})$-wave pairing 
superconductor. The distance from the (anti)vortex core is denoted by $r$.
The  intrinsic line width is taken to be $\Gamma=0.002$.}
\label{fig:ldosx2y2}
\end{figure}

Analytical solutions to Eqs.~\eqref{eq:bdgdl} can be derived in the limit $\gamma \gg 1$.
In this limit the Bogoliubov equations decouple and the solutions are given in terms of the Hankel functions
of the first and second kind, $H^{(1)}_{\alpha} (x)$ and $H^{(2)}_{\alpha} (x)$.
Thus, we make the following ansatz for the wavefunctions
\begin{eqnarray} \label{eq:ansatz1}
u(r) &= f_1(r) H_{M_+}^{(1)}(qr) + f_2(r) H_{M_+}^{(2)}(qr),
\nonumber \\
v(r) &= g_1(r) H_{M_-}^{(1)}(qr) + g_2(r) H_{M_-}^{(2)}(qr), 
\end{eqnarray}
with $q = \sqrt{2} \gamma$.
The functional form of the coefficients $f_i (r)$ and $g_i (r)$ (with $i  \in \{ 1, 2\}$) is derived in Appendix~\ref{firstAppendix}, from which it follows that  $f_2 (r) = f_1^{\ast} (r) $ and $g_2 (r) = g_1^{\ast} (r)$.
The energy spectrum of the vortex-bound sates is found to be (see Appendix~\ref{firstAppendix})~\cite{volovikJETP99,Caroli64}
\begin{equation} \label{energySpacing}
\epsilon_l  = l \left(\frac{\int_0^\infty \frac{2\sqrt{2}}{r} h^2(r')e^{2\alpha_0(r')} dr'}{\int_0^\infty e^{2\alpha_0(r')} dr'}\right) ,
\end{equation}
where $l = \frac{2\mathrm{k}-1}{2}$, with $\mathrm{k} \in \mathbb{N}$.
Hence, in agreement with the topological argument of Refs.~\onlinecite{ryuNJP10,chiu_review15,TeoPRB10,Freedman2011}, we find that 
the $(d_{x^2-y^2}+id_{xy})$-wave superconductor
does not have any zero-energy vortex-bound states [Fig.~\ref{fig:spectrum}(a)].
The first subgap state has  non-zero energy $\epsilon_{\frac{1}{2}}$
and the other low-lying  states are evenly spaced with spacing $\epsilon_1$.

\begin{figure} [t!]
 \includegraphics[width=0.45\textwidth]{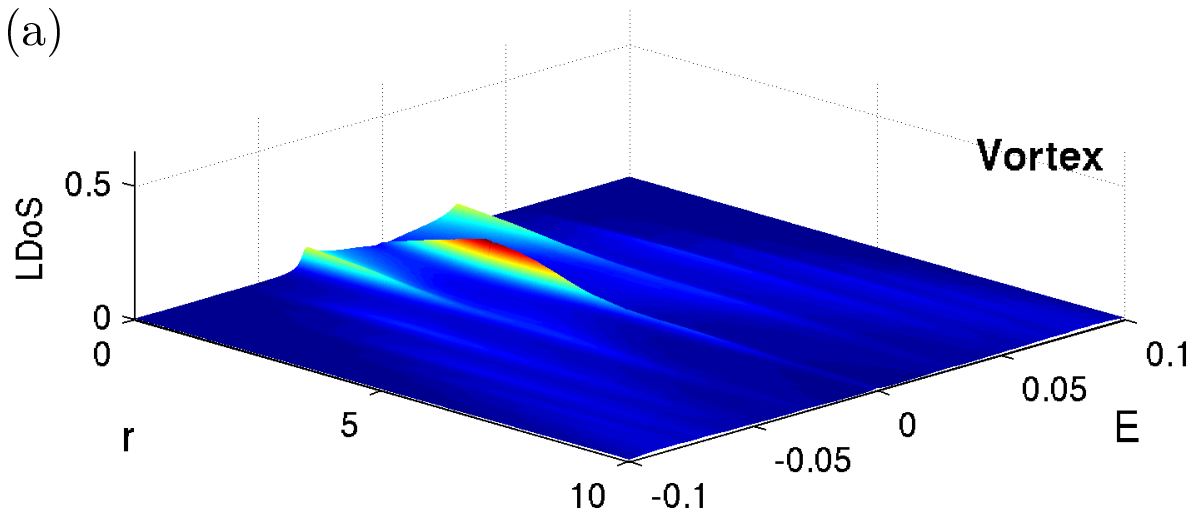}
\includegraphics[width=0.45\textwidth]{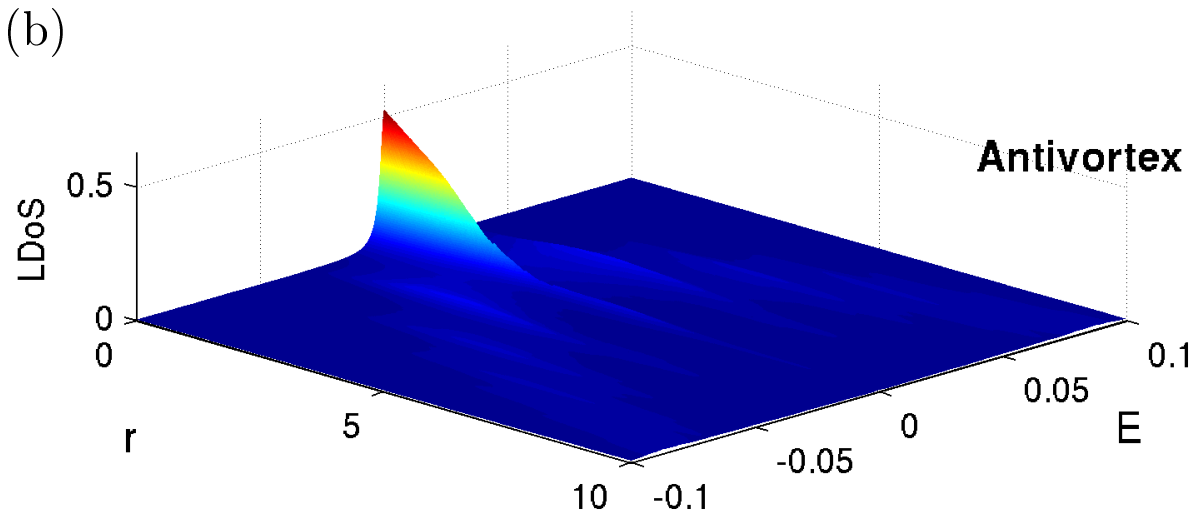}
\caption{(Color online) Local density of states of Eq.~\eqref{defLDOS} near (a) the vortex and (b) the antivortex of the $(d_{xz}+id_{yz})$-wave pairing superconductor. The intrinsic broadening is taken to be $\Gamma=0.002$.}
\label{fig:ldosxz}
\end{figure}

We observe that the bound-state energy spectrum of the vortex is the same
as the one of the anti-vortex. But there is a striking difference in
the wavefunctions between the vortex- and antivortex-bound states.
This is demonstrated in Fig.~\ref{fig:ancomp}, which plots the wavefunction amplitude
$P(r) = |u(r)|^2 + |v(r)|^2$
of the first two lowest energy vortex- and antivortex-bound states.
Indeed, from the above discussion and using the fact that 
$H^{(2)}(qr) = [H^{(1)}(qr)]^*$,
we find that
the wavefunction amplitude of the $\mathrm{k}$-th lowest energy  (anti)vortex-bound state is given by
\begin{eqnarray}
&&
P (r) = \\
&& \;
\left\{ \begin{array}{l l}
\Re [ f (r)  H^{(1)}_{\mathrm{k}+1 } (q r ) ]^2 + \Re [ g(r)   H^{(1)}_{2 - \mathrm{k}  } (q r ) ]^2  & 
\, \textrm{vortex} \\
\Re [ f (r)  H^{(1)}_{\mathrm{k} } (q r ) ]^2 + \Re [ g(r)   H^{(1)}_{1 - \mathrm{k}  } (q r ) ]^2     &  \,
\textrm{antivortex} \, .
\end{array} \right.  
\nonumber
\end{eqnarray}
We observe that  $\Re [ H^{(1)}_\alpha (r) ]$
exhibits a node at $r=0$  for all $\alpha$ except for $\alpha=0$, in which case $\Re [ H^{(1)}_0 ( 0 ) ]=1$.
Hence, it follows that for the vortex-bound states the lowest-energy wavefunction ($\mathrm{k}=1$)  is peaked at finite $r$, whereas for
the antivortex it is peaked at the origin $r=0$, see Figs.~\ref{fig:ancomp}(a) and~\ref{fig:ancomp}(b).
This finding is corroborated by our numerical simulations, which we present
in the following subsection.

\subsection{Numerical results}
\label{numerical}

To compute the energy spectrum, the wavefunction amplitudes, and the local density of states of the vortex-bound states, we discretize Hamiltonian~\eqref{Ham} in the presence of the vortex/antivortex pair~\eqref{defVortexAntivortex}
on the tetragonal lattice with $200 \times 200$ sites in the $xy$ plane and 100 points along the $k_z$ direction.
The eigenenergies $\varepsilon_{\nu, k_z}$   and eigenfunctions $\Phi_{\nu,k_z}({\bf r})$ of Eq.~\eqref{Ham} are obtained by exact diagonalization
of the disrectized Hamiltonian. Since the system has translation invariance along $z$ we can diagonalize it for each $k_z$ separately.
In the following we fix the parameters to $t=1$, $t_z=0.3$, $\mu=1.9$, and $\rho=5$. For the 
$(d_{x^2-y^2}+id_{xy})$-wave pairing state we chose
 $\Delta_0 = 0.05$, while for  the $(d_{xz}+id_{yz})$-wave state we set  $\Delta_0=0.2$.
With this parameter choice, the Fermi surface has a cigar-like shape elongated along the $k_z$ direction,
which corresponds to the shape of the Fermi surface at the $H$ point of SrPtAs~\cite{youn2011,youn2012,Biswas2013,Fischer2014}.
We have checked that different parameter values do not qualitatively change the energy spectrum and the local density of states
of the (anti)vortex-bound states. 

\begin{figure} [t!]
\includegraphics[width=0.45\textwidth]{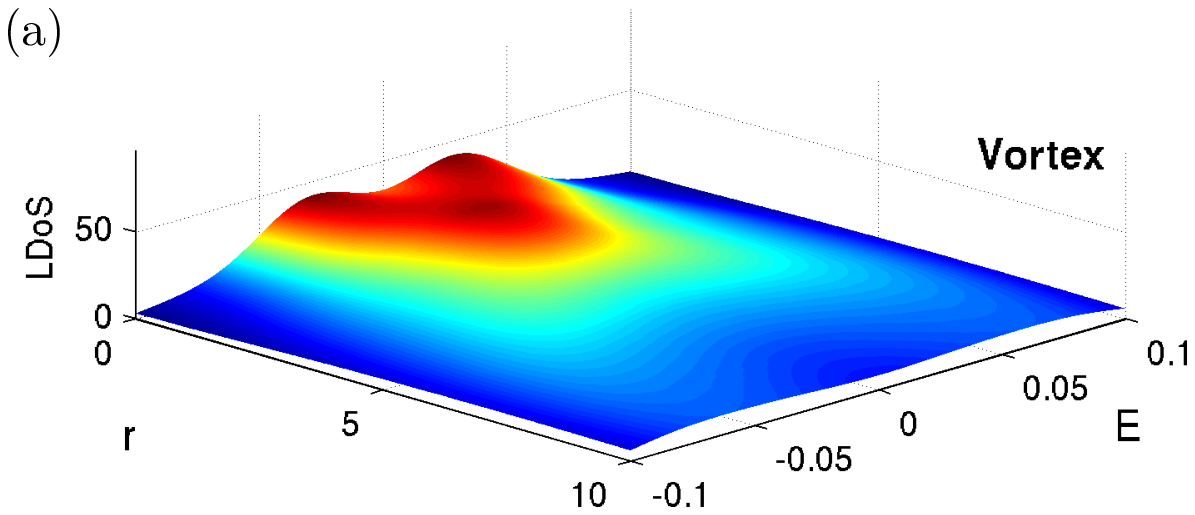}
\includegraphics[width=0.45\textwidth]{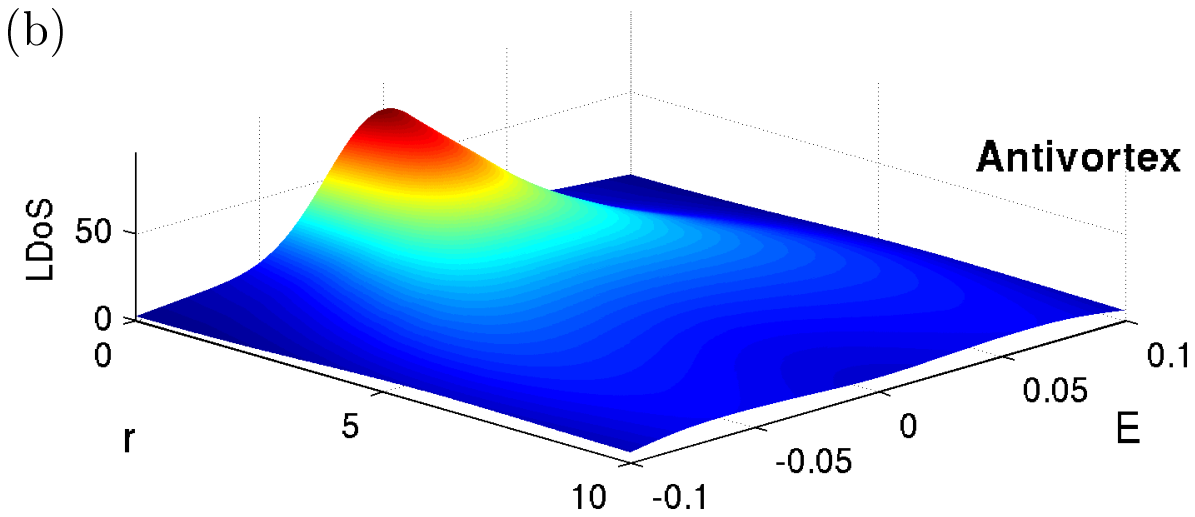}
\caption{(Color online) Broadened local density of states  (LDOS)
near the cores of (a) the vortex and (b) the antivortex of 	
the $(d_{x^2-y^2}+id_{xy})$-wave pairing superconductor. 
To simulate the effects of finite temperature and finite experimental resolution
the LDOS was convoluted with a Gaussian with FWHM $\tau=0.045$ corresponding
to the energy resolution of the experiment.}
\label{fig:cldosx2y2}
\end{figure}

 \subsubsection{Energy spectrum and wavefunction amplitude} 

In Fig.~\ref{fig:spectrum} we present the energy spectrum $\varepsilon_{\nu, k_z}$ as a function of $k_z$ of the 
$(d_{x^2-y^2}+id_{xy})$-wave and $(d_{xz}+id_{yz})$-wave pairing states. The energies of the vortex-bound states
are indicated in red, showing that vortices in the $(d_{xz}+id_{yz})$-wave state
exhibit a zero-energy flat band of bound states,
whereas vortices in the $(d_{x^2-y^2}+id_{xy})$-wave state support bound states only at finite energy.

Fig.~\ref{fig:ancomp} displays the wavefunction amplitude $\left| \Phi_{\nu,k_z}(r) \right|^2$ 
of the lowest and second lowest energy bound state at the vortex and antivortex of the 
 $(d_{x^2-y^2}+id_{xy})$-wave superconductor for $k_z=0$ .
 Our numerical results (dashed blue curves) are in excellent agreement with the analytical solutions of the (anti)vortex-bound states (solid red curves).
 As discussed in the previous subsection,  the lowest energy antivortex state exhibits a peak at $r=0$, whereas the vortex state peaks at a non-zero $r$. 
 For the second lowest energy bound state the behavior is opposite: The vortex state is peaked a the origin, while the maximum of the antivortex state 
 is at finite $r$. We note that these trends are independent on the $k_z$ value, since  the overall shape of the wavefunctions is given by the order $k$ of the 
Hankel functions $H^{(1)}_k$, which only depends on the vorticity $n$ and the quantum number $l$ (see Sec.~\ref{analytical}).

We remark that there is a similar asymmetry between the vortex- and antivortex-bound states of the 
$(d_{xz}+id_{yz})$-wave pairing superconductor. That is, the zero-energy anti\-vortex-bound state
has a maximum at $r=0$, while the zero-energy vortex-bound state is peaked at finite $r$. 
Again, this is a consequence of the difference in the order $k$ of the Hankel functions $H^{(1)}_{k}$ describing the
bound-states of the (anti)vortex (cf.~discussion in Refs.~\onlinecite{volovikJETP99,krausPRL08,krausPRB09,moellerPRB11}).

 \subsubsection{Local density of states} 

\begin{figure} [t!]
\includegraphics[width=0.45\textwidth]{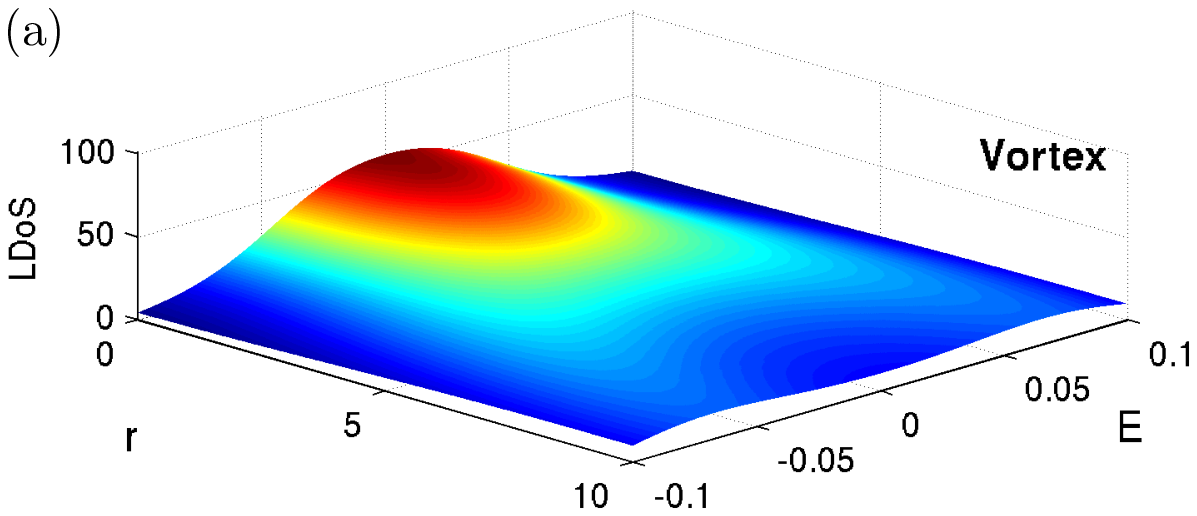}
\includegraphics[width=0.45\textwidth]{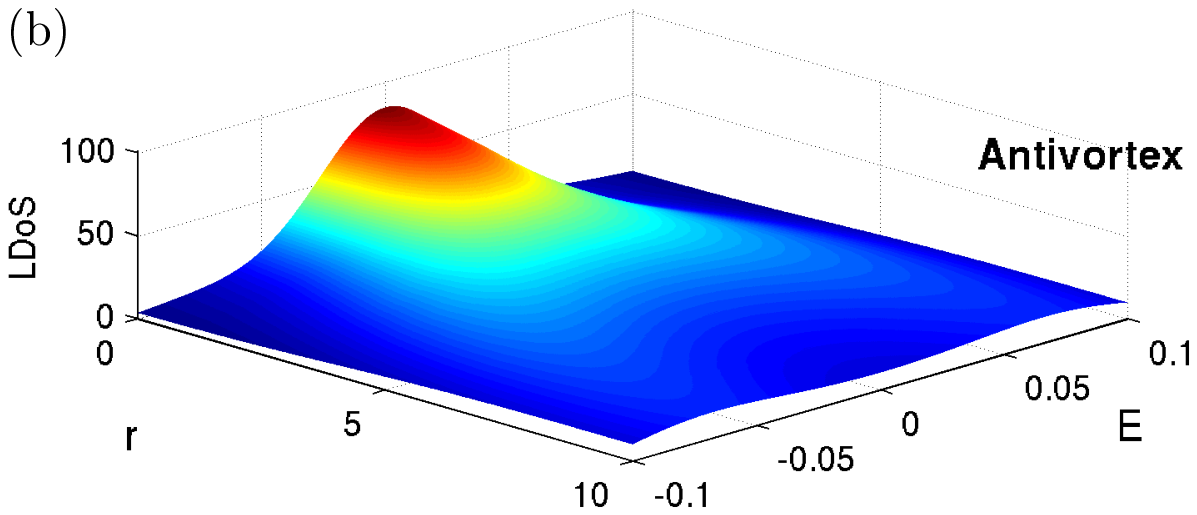}
\caption{(Color online)  Broadened local density of states (LDOS) near the core of (a) the vortex and (b)
the antivortex of the $(d_{xz}+id_{yz})$-wave pairing state. The LDOS was convoluted with a Gaussian with FWHM $\tau=0.061$ corresponding to the energy resolution of the experiment. }
\label{fig:cldosxz}
\end{figure}
 
The vortex-bound states of type-II superconductors can be probed by scanning tunneling spectroscopy of the surface density of states~\cite{hessPRL89,hessPRL90,sunWahlNatComm15}.
 To facilitate direct comparison with experimental measurements, we calculate the local density of states (LDOS) around
 the vortex and antivortex cores.
 The local density of states as a function of distance $r$ from the vortex (or antivortex) center is given by
\begin{align} \label{defLDOS}
	\rho( E ,r) = \frac{-1}{N}\frac{1}{4\pi} \Im \left(\sum_{k_z} \sum_\nu \frac{[\Phi_{\nu,k_z}(r)]^\dagger \Phi_{\nu,k_z}(r)}{E + i\Gamma - \epsilon_{\nu,k_z}}\right) ,
\end{align}
where $\nu$ labels the eigenstates $\Phi_{\nu,k_z}$ and eigenvalues $\epsilon_{\nu,k_z}$,
$E$ denotes the energy, and $\Gamma$ represents an intrinsic broadening due to disorder.

Figures \ref{fig:ldosx2y2} and~\ref{fig:ldosxz} show the LDOS near the core of the vortex/antivortex 
of the  $(d_{x^2-y^2}+id_{xy})$-wave and the $(d_{xz}+id_{yz})$-wave states, respectively.
The bound states appear as sharp peaks with energy spacing $\epsilon_1$, given by Eq.~\eqref{energySpacing}.
Comparing Fig.~\ref{fig:ancomp} with Figs.~\ref{fig:ldosx2y2} and~\ref{fig:ldosxz}, we find that the non-zero $k_z$ dispersion
of the finite-energy bound states leads to a small  broadening in energy of the LDOS peaks. 
The Majorana flat-band states of the $(d_{xz}+id_{yz})$-wave paring superconductor, on the other hand, 
have no $k_z$ dispersion and therefore give rise to   
very sharp zero-energy peaks  near the center of the vortex and antivortex cores, see Fig.~\ref{fig:ldosxz}.
Importantly, the asymmetry between the vortex and antivortex-bound states
is directly visible in the local-density of states:
 The lowest-energy antivortex-bound sta\-tes are peaked at $r=0$, while  
 the lowest-energy bound states of the vortex have nodes at $r=0$. 
 The reason for this distinction was discussed in Sec.~\ref{analytical}.

In a scanning tunneling scpectrocopy experiment the LDOS is smeared
by temperature broadening~\cite{hessPRL89,hessPRL90,sunWahlNatComm15,maeno_physica_C}. To simulate this we convolute the LDOS with a 
Gaussian with full width at half-maximum $\tau$ corresponding to the experimental energy resolution. We choose $\tau$ to be of the order of
two times the level spacing of the bound states $\epsilon_1$, Eq.~\eqref{energySpacing}. Figures~\ref{fig:cldosx2y2} and~\ref{fig:cldosxz} show the broadened LDOS for the two pairing symmetries.
We observe that the LDOS peak of the vortex is much broader than that of the antivortex. Moreover, 
the peak of the vortex is about half the height of that of the antivortex and 
it exhibits two ridges which disperse away to larger $r$.  This is because the height of the LDOS peak is determined by the broadening of
 the lowest-energy state, while 
 for the vortex it is due to the broadening of several low-energy states.
In conclusion, we find that the asymmetry between the vortex and the antivortex can be detected
in  the LDOS even at temperatures $T$ larger than the level spacing
$\epsilon_1$.

\section{Discussion and Final Remarks}
\label{sec:summary}

In this paper we have used large-scale exact diagonalization and analytical methods to study  
the structure of vortex-bound states in chiral $d$-wave superconductors. 
We have shown that vortices in the chiral \mbox{($d_{xz} \pm id_{yz}$)}-wave state bind
dispersionless zero-energy states, which form a doubly degenerate Majorana flat band (Fig.~\ref{fig:spectrum}).
The stability and robustness of these zero-energy vortex-bound states is guaranteed by 
a Chern-Simons topological invariant.
For the ($d_{x^2-y^2} \pm i d_{xy}$)-wave superconductor we found that vortex-bound states
exist only at finite energy.  We have computed the LDOS
near the core of the vortex and antivortex of these chiral $d$-wave superconductors. 
Importantly, we found a pronounced asymmetry in the LDOS between the vortex and the antivortex:
The lowest-energy peak in the LDOS of the antivortex has its maximum at $r=0$, while the lowest-energy peak of the vortex is centered at $r \ne 0$ (Figs.~\ref{fig:ldosx2y2} and~\ref{fig:ldosxz}).
Moreover, we have shown that the Majorana vortex-bound states of the \mbox{($d_{xz} \pm id_{yz}$)}-wave
superconductor give rise to a particularly sharp peak in the LDOS, since these zero-energy bound states do not exhibit any dispersion in energy (Fig.~\ref{fig:ldosxz}).

The asymmetry in the LDOS between the vortex and the antivortex can in principle be used as a clear experimental fingerprint of the chiral order parameter symmetry. In practice, however, 
this might naively only be possible at temperatures $T$ smaller than the
level spacing $\epsilon_1$ between the bound states, since the LDOS is smeared by temperature broadening.
The energy spacing $\epsilon_1$  [cf.~Eq.~\eqref{energySpacing}] is of the order of
$\Delta_0^2 / E_F$, where $\Delta_0$ is the superconducting gap amplitude and $E_F$ is the Fermi energy.
For a typical unconventional superconductor this corresponds to a temperature of about $\sim100 \mu$K,
which is below the reachable temperature regime of current state-of-the-art STM machines~\cite{singhSTM}.
However, a clear asymmetry between the LDOS of the vortex and the antivortex remains
even at temperatures $T$ of the order of $\epsilon_1  < T  < \Delta_0$, see Figs.~\ref{fig:cldosx2y2} and~\ref{fig:cldosxz}. That is, 
even though the individual LDOS peaks of the bound states cannot be resolved at
a temperature $T > \epsilon_1$, the broadened peak around the vortex 
is much wider and about half as high as the one of the antivortex.
Hence, we believe that the predicted asymmetry is
experimentally accessible for realistic materials, such as URu$_2$Si$_2$~\cite{Hsu14,Kasahara07,Li13,Sumiyoshi14,Schemm2014,Mydosh2011} and SrPtAs~\cite{youn2011,youn2012,Biswas2013,Fischer2014}, and hope
that our findings will stimulate future STM experiments on these interesting unconventional 
superconductors.

\acknowledgments
The authors thank M.~Sigrist and P.~Wahl for useful discussions.
This work was supported by the Max Planck-UBC Centre for Quantum Materials.

\appendix

\section{Derivation of the vortex-bound states}
\label{firstAppendix}

In this appendix, we derive analytical formulas for 
the solutions to the BdG equations~\eqref{eq:bdgdl}
in the limit $\gamma \gg 1$.
For brevity, we restrict our discussion to the ansatz
\begin{eqnarray} \label{ansatzApp}
u(r) &= f_1 (r) H_{M_+}^{(1)}(qr),
\quad
v(r) &= g_1(r) H_{M_-}^{(1)}(qr) 
\end{eqnarray}
for the wavefunctions.
 The solutions for the ansatz in terms of the Hankel functions of the second kind 
 $H_\alpha^{(2)}$ can be derived in an analogous manner [cf.~discussion above Eq.~\eqref{someIntegral}].
Assuming $ r \gg 1/q$, we can approximate
the Hankel function $H^{(1)}_{\alpha}$ by~\cite{Abramowitz}
\begin{eqnarray} \label{asymptoticHankel}
		H^{(1)}_\alpha(qr) \approx \frac{\exp[  i (qr - \alpha \pi/2)]}{\sqrt{qr}} .
\end{eqnarray}
We observe that the asymptotic form of $H^{(1)}_{M_-}$ is proportional to $H^{(1)}_{M_+}$ times a phase factor, i.e.,
 $H^{(1)}_{M_-} = (-1)^{\mathrm{k}} i H^{(1)}_{M_+}$, since
$M_- - M_+ = - 2l$  and $l = \frac{2\mathrm{k}-1}{2}$. 
In addition, we find that the derivatives of the asymptotic Hankel function \eqref{asymptoticHankel}  are given by
\begin{align}
\frac{dH^{(1)}_{M_+}}{dr} &= \left(iq - \frac{1}{2r}\right)H^{(1)}_{M_+} , \\
\frac{d^2H^{(1)}_{M_+}}{dr^2} &= \left(-q^2 - \frac{3}{4r^2} - \frac{iq}{r}\right)H^{(1)}_{M_+} .
\end{align}
Inserting ansatz~\eqref{ansatzApp} into Eqs.~\eqref{eq:bdgdl} and using the above approximations
yields the following differential equations for $f_1$ and $g_1$ in the limit $q r \gg 1$
\begin{subequations} \label{eq:bdgf1}
\begin{eqnarray} 
\left[ -\frac{1}{2} \widetilde{L_{M_+}} - \gamma^2\right]f_1 + \frac{i (-1)^{\mathrm{k}} }{\gamma}\widetilde{D_+} g_1 & = \epsilon f_1 , \\
\left[ \frac{1}{2} \widetilde{L_{M_-}} + \gamma^2\right]g_1 + \frac{i (-1)^{\mathrm{k}+1} }{\gamma}\widetilde{D_-} f_1 & =  \epsilon g_1 ,
\end{eqnarray}
with the differential operators
\begin{eqnarray}
\widetilde{L_\pm } 
= \partial_r^2 + 2iq\partial_r - \frac{1}{4r^2} \left[ 5+4 ( M_\pm) ^2+4q^2r^2 \right]
\end{eqnarray}
and  
\begin{eqnarray}
&& \widetilde{D_\pm} =  \frac{h^2}{r^2}\left(\frac{9}{4}\mp l-l^2 \right) + 2hh'\left(\frac{\pm l}{r}-iq\right)  \\
&& \quad +h^2q\left(\frac{\pm 2il}{r}+q\right) + \frac{2h}{r}\left[ h \left(l-iqr \right) - h'r\right] \partial_r - h^2 \partial_r^2 .
\nonumber
\end{eqnarray}
\end{subequations}
The  set of equations~\eqref{eq:bdgf1} can  be analyzed in a perturbative approach. 
In the small $q$ limit and focusing on solutions that are decaying as $ r \to \infty$,
we find that at the first order in $q$ the equations are solved by the exponential functions
\begin{subequations} \label{ansatz_f1_g1}
\begin{eqnarray}
	f_1(r) & =& \exp\left([\alpha_0(r) + i\beta_0(r)] + \frac{i}{q}[\alpha_1(r) + \beta_1(r)]\right) \label{eq:f1ansatz}, \; \; \; \; \;  \; \; \;  \\
	g_1(r) & =&\exp\left([\alpha_0(r) - i\beta_0(r)] + \frac{i}{q}[\alpha_1(r) - \beta_1(r)]\right) \label{eq:g1ansatz} , \;
\end{eqnarray}
with
\begin{eqnarray}
	\alpha_0 &=& -\int_0^r \sqrt{2} h^2(r') dr'  \quad
	\textrm{and} \quad
	\beta_0  = \frac{\mathrm{k}\pi}{2} .
\end{eqnarray}
\end{subequations}
The functions $\alpha_1$ and $\beta_1$ in Eqs.~\eqref{ansatz_f1_g1} describe corrections at the next order in $q$ and can be determined
by approximating $f_1$ and $g_1$ by
\begin{eqnarray} \label{eqF1G1}
	f_1(r) & \approx & e^{\alpha_0} \left(1 + \frac{i}{q}\left(\alpha_1(r) + \beta_1(r)e^{-2\alpha_0}\right)\right) , \\
	g_1(r) & \approx & (-1)^{\mathrm{k}+1}e^{\alpha_0} \left(1 + \frac{i}{q}\left(\alpha_1(r) - \beta_1(r)e^{-2\alpha_0}\right)\right)  ,
	\nonumber
\end{eqnarray}
and substituting this ansatz into Eq.~\eqref{eq:bdgf1}. 
Equating terms which are $q$-independent and solving the resulting differential equations for $\alpha_1$ and $\beta_1$, we obtain
\begin{align} \label{solAlpha1Beta1}
	\alpha_1(r) &= - \int_0^r 3h^4(r') - \sqrt{2} h(r')h'(r') dr' , \cr
	\beta_1(r) &= - \int_r^\infty \left(\epsilon-\frac{2\sqrt{2}l}{r'} h^2(r')\right)e^{2\alpha_0(r')} dr' .
\end{align}
We observe that the solutions $f_1$ and $g_1$, Eq.~\eqref{ansatz_f1_g1}, with $\alpha_1$ and $\beta_1$ given by Eq.~\eqref{solAlpha1Beta1},
are well behaved for large $r$ since the radial vortex profile $h(r)$ approaches $1$ at large distances.

The coefficients $f_2$ and $g_2$ for the Hankel functions of the second kind $H_\alpha^{(2)}(x)$ in Eq.~\eqref{eq:ansatz1} can be derived in a similar manner, repeating the same steps as above. We find $f_2(r) = f_1^*(r)$ and $ g_2(r) = g_1^*(r)$. 
Finally, we are ready to construct the full solution to the differential equations~\eqref{eq:bdgdl}, which is given in terms
of a superposition of $H_\alpha^{(1)}$ and $H_\alpha^{(2)}$. 
The full solution needs to be regular at the origin $r=0$, which leads to the condition that
$\Im [ f_1(0)] = \Im [ g_1(0)] = 0$. That is, $f_i(0)$ and $g_i(0)$ need to be the same for the two Hankel functions, 
such that the imaginary singular part of the Hankel function is eliminated at the origin.
From Eq.~\eqref{eqF1G1} we find that this requirement is equivalent to $\alpha_1(0) = \beta_1(0) = 0$. 
The condition for $\alpha_1$ is automatically satisfied; the one for $\beta_1$, however, yields
\begin{eqnarray} \label{someIntegral}
 \int_0^\infty \left(\epsilon-\frac{2\sqrt{2}l}{r} h^2(r )\right)e^{2\alpha_0(r)} dr  = 0 ,
 \end{eqnarray}
 which determines the energy spectrum of the vortex-bound sates $\epsilon_l$,
 which is given in Eq.~\eqref{energySpacing}~\cite{volovikJETP99,Caroli64}.

\bibliography{refs_v2}

\begin{thebibliography}{50}
\bibitem[{foo({\natexlab{a}})}]{footnote1}
\bibinfo{note}{This is done by minimizing a combination of two errors: the
  difference in angle between the corrected phase and the original phase (when
  $C(y)=1$), and the difference in angle of the corrected phase at the two
  boundaries.}
\bibitem[{foo({\natexlab{b}})}]{footnote2}
\bibinfo{note}{Because the continuum model is obtained by taking the lowest
  order Taylor series terms from the lattice model, we get $\cos(k)\approx
  1-k^2$ and $ \sin(k) \approx k$. Thus, the $k^2$ terms will flip sign,
  whereas the $k$ terms will not, causing a flip in chirality when relating the
  lattice model to the continuum model.}
\end{thebibliography}

 \end{document}